\documentclass[aps,prl,preprintnumbers,twocolumn,groupedaddress,footinbib]{revtex4-1}

\pdfoutput=1

\usepackage{amsmath,amssymb,mathrsfs}
\usepackage{slashed}
\usepackage{xcolor}
\usepackage[colorlinks=true,linkcolor=blue,citecolor=blue,urlcolor=violet]{hyperref}
\usepackage{tikzfeynman}
\usepackage{soul}
\usepackage{ulem}

\newcommand{\GeV}{\mathrm{GeV}}

\newcommand{\be}{\begin{equation}}
\newcommand{\ee}{\end{equation}}

\definecolor{viola}{RGB}{134,41,198}

\begin{document}

\title{Indirect Detection  of  Composite (Asymmetric) Dark Matter} 

\author{Rakhi Mahbubani}
\email{rakhi@cern.ch}
\affiliation{Albert Einstein Center for Fundamental Physics, Institute for Theoretical Physics, University of Bern, Sidlerstrasse 5, CH-3012 Bern, Switzerland}
\affiliation{Theoretical Physics Department, CERN, 1211 Geneva 23, Switzerland}

\author{Michele Redi}
\email{michele.redi@fi.infn.it}

\author{Andrea Tesi}
\email{andrea.tesi@fi.infn.it}
\affiliation{
INFN Sezione di Firenze, Via G. Sansone 1, I-50019 Sesto Fiorentino, Italy
}
\affiliation{Department of Physics and Astronomy, University of Florence, Italy
} 

\preprint{CERN-TH-2019-125}

\begin{abstract}
Dark Matter can form bound states upon the emission of quanta of
energy equal to the binding energy. The rate of this process is large for strongly-interacting Dark Matter, and further enhanced by long distance effects.  The resulting monochromatic
and diffuse $\gamma$-rays can be tested in indirect detection
experiments. If Dark Matter has
electroweak charge, indirect signals include multiple observable photon lines for masses in the TeV range. Else if it couples only via a dark photon
portal, diffuse spectra from dwarf galaxies and CMB reionization set
powerful limits for masses below a TeV. This mechanism provides a powerful
means of probing Asymmetric Dark Matter today. 
\end{abstract}

\maketitle

\paragraph{\bf Introduction.}\,\,

Having no annihilation rate today, Asymmetric Dark Matter (DM) is
largely untestable in indirect detection experiments
absent some non-minimal assumption (e.g a remnant annihilating
component, or decaying relic, see \cite{review,admreview} for reviews
on the subject).  We argue in a compelling analogy with Standard Model
(SM) baryons that `nucleons' of a dark strong sector
naturally emit a light particle on forming bound state `nuclei'.  The 
rate for this process is calculable semi-analytically in
the limit of shallow bound states, and can be large, allowing us to probe dark nuclear
Asymmetric DM in existing indirect-detection experiments.
(See
\cite{Pearce:2013ola,Pearce:2015zca,Detmold:2014qqa,Mitridate:2017izz,Baldes:2017gzu}
for related work).  

While our considerations apply more generally, \textit{e.g.}  to
conventional thermal DM, we focus here
on models where DM is asymmetric, and composite due to dark strong interactions, in
close analogy with SM nucleons.  The thermal abundance of DM is reproduced for masses around 100 TeV \cite{Antipin:2015xia} 
so that for masses below this value the symmetric component of DM is subleading.
Composite DM can be simply realised as
the lightest baryon in an SU($N$) confining gauge theory with dark fermions
that are vectorial under the SM \cite{Antipin:2015xia}, see \cite{Kribs:2016cew} for a review. DM
cosmological stability follows from the accidental dark baryon-number
conservation, which also guarantees the stability of the lightest 
state in each baryonic sector. The dark sector is roughly characterised at low energies by
$i)$ the mass of the lightest dark baryon, $M$, which constitutes the
DM; $ii)$ the mass of the `dark pion', $M_\pi \lesssim M$, that sets
the typical range for nuclear interactions amongst the baryons (we
assume that this state is cosmologically unstable); $iii)$ the mass of a weakly-coupled mediator
external to the strong sector, $M_V$, \textit{e.g.} SM gauge
bosons. We will assume that the spectrum features nuclear bound states
with binding energies $E_B > M_V$, focussing in particular on the
nucleus with baryon number 2, `dark deuterium'.

\begin{figure}
\centering
\begin{tikzpicture}[line width=1.5 pt, scale=1.7]
	\node at (-1.7,0.35) {DM};
	\node at (-1.7,-0.35) {DM};
	\draw[] (-1.5,0.35)--(0,0);
	\draw[] (-1.5,-0.35)--(0,0);
	\draw[line width=3pt, color=black] (0.7,-0.3)--(0,0);
	\draw[vector,color= viola] (0,0)--(0.7,.3);
	\draw[vector,color= viola] (-1.38,-0.31)--(-1.38,0.31);
	\draw[vector,color= viola] (-1.18,-0.26)--(-1.18,0.26);
	\draw[dotted,color= viola] (-1.05,0)--(-0.86,0);
	\draw[vector,color= viola] (-0.78,-0.17)--(-0.78,0.17);
	\draw[fill, color=gray] (0,0) circle (.3cm);
	\node at (1,-0.33) {$^2$DM$^*$};
	\node at (1,.33) {~~~~~~$\gamma/Z/W/\gamma_D$};
\end{tikzpicture}
\vspace{-0.2cm}
\caption{
\label{fig:Feynprocess}\it
Bound state formation considered in this letter. The process is affected by (long-) short-distance (`weak') `nuclear'  physics. 
The bound state can be unstable and decay to the ground state emitting
additional lines.}
\end{figure}
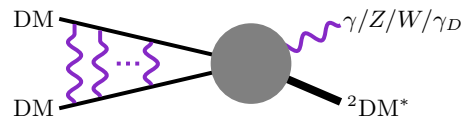

The cosmological production of dark nuclei was studied in
\cite{Redi:2018muu}; here we consider observational implications.
At DM velocities relevant for indirect detection, dark deuterium is produced essentially at rest through
emission of a quantum of energy $E_B$. The process is depicted in
Fig.~\ref{fig:Feynprocess}.  We will consider two main
scenarios, characterised by the properties of the mediator carrying
the quantum of energy emitted in bound state formation. The first is
automatically realised if DM has electroweak charges, and gives rise
to emission of SM gauge bosons, in particular to monochromatic
photons.  These are constrained 
by observations of the Galactic centre by FERMI and HESS. In the second
scenario DM is neutral under the SM but charged under an additional
(broken) U(1). Dark deuterium is then produced in association with a dark photon that later decays to SM particles.
The resulting diffuse photon signal can be tested using observations
of dwarf spheroidal galaxies, and CMB re-ionization. As we will
discuss shortly, both alternatives can be analysed by exploiting the analogy with deuterium production in the SM, yet they present very different experimental signatures.

\paragraph{\bf Nuclear Cross sections.}\,\,
We consider the production of shallow nuclear bound states with $E_B
\ll M$. As shown by Bethe and Longmire \cite{Bethe:1950jm}, and more recently 
derived systematically using nucleon effective field theories
\cite{Kaplan:1998tg}, at low energy the cross section for formation of a shallow bound state
does not depend on the details of the potential, but simply on the
parameters of the effective range expansion. The amplitude for
elastic scattering is determined by the phase shift $\delta$, which for 
$s$-wave scattering admits the following expansion,
\be
\mathcal{A}=\frac{4\pi}{M} \frac{1}{p\cot\delta -i p}\,,\quad p \cot \delta=-\frac 1 a + \frac 1 2 r_0 p^2+ \dots\,,
\ee
where $p$ is the momentum of the incoming states in the centre-of-mass frame, $a$ is the
scattering length and $r_0$ is the range of the interaction, $r_0\sim 1/M_\pi$.
Importantly for a shallow $s$-wave bound state the scattering length is
determined by the binding energy, $1/a\approx  \sqrt{E_B M}$, as can
be seen from the pole of $\mathcal{A}$. The deuteron formation cross section is computed in
terms of the binding energies/scattering lengths of the $np$ elastic
scattering channels $^1 S_0$ and $^3 S_1$
\cite{Savage:1998ae}. Indeed in the presence of bound states
(\textit{i.e.} poles of the elastic amplitude for imaginary momenta
$p$),  the elastic scattering amplitude allows one to extract the coupling of the deuteron to two nucleons from
 the residue at the pole  and calculate the cross section with Feynman diagrams, see \cite{Kaplan:2005es} for a review.

In this letter we focus on dark nuclear transitions induced by magnetic dipole interactions,
\begin{equation}\label{magnetic-dipole}
\kappa \frac { e} {M} N^\dagger J_3 \,(\vec\sigma\cdot \vec B)\, N\,,
\end{equation}
where $N$ is the non-relativistic dark nucleon field, $J_3$ is the third
component of isospin, and $\kappa \sim 1$ for strongly coupled nucleons. 
This interaction induces transitions with selection rules $\Delta L=0$ and $\Delta S=1$, allowing for bound state formation from an initial $s$-wave state.
Nuclei can also be produced through electric dipole transitions but this process is typically suppressed at low velocities, see Ref. \cite{future}.

The  cross section for the formation of an $s$-wave bound state through
a magnetic transition reads \cite{Redi:2018muu}
\begin{equation}\label{magnetic-xsec}
(\sigma v_{\rm rel})_{NN \to D\gamma}^{\rm mag}\approx \kappa^2 K_M  \frac  {4\pi  \alpha}{M^2} \bigg(\frac{E_{B_f}}{M}\bigg)^{\frac 3 2} \bigg(1-\frac{a_i}{a_f}\bigg)^2,
\end{equation}
where $a_{i,f}$ are the scattering lengths of initial and final state,
$v_{\rm rel}$ is the relative velocity of the incoming states in the
centre-of-mass frame,
and $K_M$ is a group theory factor, equal to 1 for deuteron formation
in the SM.  

This cross section can be significantly modified by long-distance
effects due to forces external to the strong sector
\cite{Hisano:2006nn}. Such long-distance
modification is intimately tied to the mechanism of bound-state formation,
which can only take place through emission of the light quanta
that
are responsible for the effect. For electroweak constituents these forces are just SM gauge
interactions, while for SM-neutral constituents we consider the possibility that they are associated
with exchange of a dark photon. In the case of DM annihilation the long-distance effects can be factorised so that  $\sigma \approx {\rm SE} \times \sigma_{\rm short}$ where 
SE  is the Sommerfeld enhancement factor that takes into account the
distortion of the initial wave-function due to long-range forces, and
$\sigma_{\rm short}$ is the perturbative cross section.
As we will discuss in detail in \cite{future}, for bound-state
formation the long distance effects often cannot be simply factorised.
As a result, in this letter we carry out a full quantum-mechanical
computation of the cross section by explicit solution of the
Schroedinger equation to obtain the physical wave-function, see \cite{Asadi:2016ybp}.

The reduced wave-function describing $s$-wave scattering of two
DM particles of mass $M$,
in a given spin/charge sector, $u(r)= \sqrt{4\pi} r \psi(r)$ satisfies the radial
Schroedinger equation,
\begin{equation}\label{equation}
-\frac {1}{M} \frac {d^2 u}{dr^2} + V(r) \,u = E \,u\,,
\end{equation}
where $E=M \beta^2$ for $\beta=v_{\rm rel}/2$. The wave-function $u(r)$ is in general a vector, on which we impose physical
boundary conditions: $u(0)=0, \, u'(r_\infty)- i p u(r_\infty)=\sqrt{4\pi} e^{-i p r_\infty} u_0 $ where $u_0$ denotes the 
DM initial state.

The potential $V(r)$, defined in a given spin/charge sector, contains a long-distance
part, associated for example with electroweak interactions, as well as a
short-distance, spherically symmetric nuclear potential $V_N$
that respects the flavour symmetry of the strong dynamics.
To leading order the nuclear potential must simply reproduce the
correct binding energies and range of interaction. We choose to
parametrise it using a 
spherical well in each irreducible representation $a$ of the global symmetry of the nuclear interactions, $V^N_a =- V_a \theta(r_0-r)$.
The depth of the well $V_a$ determines the binding energy, which we
select in order to have a single shallow bound state per
channel \footnote{An approximate formula for the relation between the
  depth of a spherical well and the binding energy is 
$V_a/M\approx \pi^2/(4 r_0^2 M^2)+ 2 \sqrt{E_{B_a}/M}/(r_0 M)$}.
The reduced wave-function describing the corresponding bound state is known
analytically in the isospin-symmetric limit, and is given roughly by $u_f(r) \approx (4 M E_{B_f})^{\frac 1 4} \exp(-\sqrt{M
  E_{B_f}}r)$. The magnetic cross section can then be computed as follows
\begin{equation}
(\sigma v_{\rm rel})^{\rm mag}= 8 \kappa^2\, \alpha \frac{E_{B_f}^3}{M^2}\times
\bigg|\int dr\, u_i^\dagger J_3 u_{f}\bigg|^2\,,
\label{eq:numerical}
\end{equation}
where $u_i$ and $u_f$ are the reduced wave-functions of initial and final states. We take $\kappa=1$ for the remainder of this letter.

\paragraph{\bf Composite SU(2)-triplet DM}\,\,

We consider a scenario with a fermionic dark
nucleon $V$ that transforms as a triplet of SU(2)$_L$.
This can be realised in an SU(3) dark gauge theory with 3 flavours
\cite{Antipin:2015xia}.
Like for the wino, electroweak symmetry-breaking effects induce a
mass splitting $\Delta=165$  MeV between the charged $V_\pm$ and neutral
$V_0$ components.
Collider bounds due to dark pion production require $M\gtrsim $ TeV.

The  nuclear potential being SU(2)$_L$ symmetric, all
composite states can be classified according to their spin and weak isospin
in each partial wave.
The lightest dark nuclei (isotopes of dark deuterium) are $s$-wave
  bound states of two dark nucleons $V$, 
with isospin-spin $3\times 3=1_0 + 3_1 + 5_0$; we name them $D_1,D_3$ and
$D_5$ respectively.  The selection rules of the magnetic-dipole operator in
eq.~\eqref{magnetic-dipole} allow for $s$-wave transitions in isospin channels
$1_0\leftrightarrow 3_1$ and $3_1\leftrightarrow 5_0$.  
Cosmological production of bound
states being typically small for DM masses in the TeV
range \cite{Redi:2018muu}, we will take DM today to be
composed entirely of neutral nucleons, $V_0$. Anti-symmetry of its wave-function implies that an $s$-wave initial
state must have spin 0.  The magnetic dipole transition then allows for
the production of an $s$-wave spin-1 nuclear bound state, the neutral component of the
SU(2)$_L$ triplet, $D_3^0$.   Including for simplicity only the singlet nuclear potential and electro-weak interactions,
the potential in the charg- 0 spin-0 subsector containing $V_+ V_-$ and $V_0 V_0$ reads
\begin{equation}
V_{Q=0}^{S=0}=  \left( \begin{array}{cc} 2\Delta - A & -\sqrt{2} B \\
                          -\sqrt{2} B & 0 \end{array}\right)+
                      V^N_{1}(r)  \left( \begin{array}{cc} \frac{2}{3}
                                           & \frac{\sqrt{2}}{3} \\
                                           \frac{\sqrt{2}}{3} &
                                                                \frac{1}{3} \end{array}\right),
\end{equation}
where $A = \alpha/r + \alpha_2 c_W^2 e^{-M_Z r}/r$, $B = \alpha_2
e^{-M_W r}/r$ are the usual electroweak contributions while $V^N_I(r)$
is the nuclear potential in the isospin-$I$ channel,  rotated to the
charge basis.
In the spin-1 channel we neglect
small corrections due to electroweak effects and take
$V_{Q=0}^{S=1}=V^N_{3}(r)$.  

\begin{figure}[t!]
\centering
\includegraphics[width=.98\linewidth]{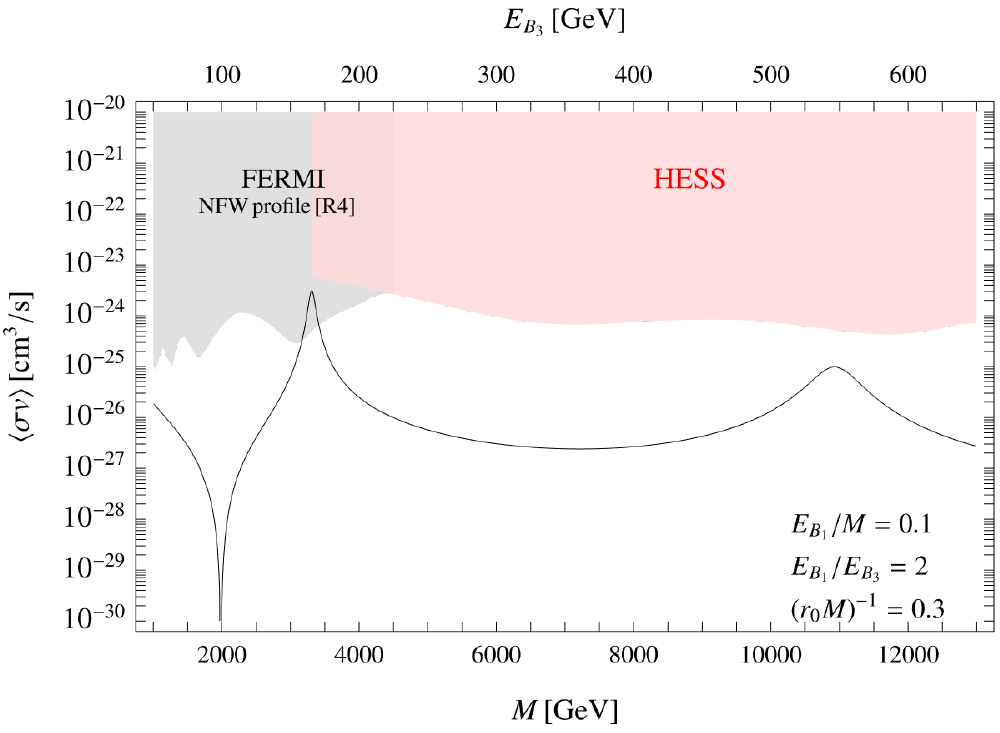}
\vspace{-0.2cm}
\caption{
\label{fig:triplet-indirect}
\it Bound state production cross section for a composite fermion
triplet of SU(2) by photon emission, $V_0V_0\to D_3^0\gamma$.
We assume negligible potential in the quintuplet channel. The
gray (red) region is the exclusion due to $\gamma$-ray lines from
FERMI \cite{FERMI-LAT} (HESS \cite{HESS2018}) in our galactic centre.}
\end{figure}

In Fig. \ref{fig:triplet-indirect} we report the cross section for
the
production of $D_3^0$ by emission of a photon, and compare with experimental constraints on $\gamma$-ray line spectra in the
galactic centre due to FERMI \cite{FERMI-LAT}, and HESS (as extracted
from \cite{HESS2018}). In making the comparison we need to account for
the reduced energy of a photon coming from bound-state formation, as
compared with that from direct annihilation of DM to photons; while the DM number density
is determined by $M$, the energy emitted is $E_B\ll M$. Thus to
extract experimental bounds the limit on the annihilation cross
section must be rescaled as follows
\begin{equation}
\langle \sigma_{D\gamma} v_{\rm rel}\rangle < 2 \left( \frac {M}{E_B}\right)^2   \langle \sigma_{\gamma \gamma} v_{\rm rel} \rangle_{M_{\rm DM}=E_B}\,,
\label{eq:rescaling}
\end{equation}
where the factor of 2 is due to the emission of a single photon in
bound state formation.

For a conservative choice of parameters, formation of dark deuterium
produces a signal just within FERMI sensitivity for masses around 3-4
TeV, where the cross section is significantly enhanced due to the
presence of a virtual zero-energy resonance in the initial state channel.  Note that the position of the peak is slightly shifted with
respect to that seen in annihilation of wino DM due to the strong nuclear potential, which
induces a shift in the binding energy of the shallow bound state.  The dip in the
cross section around 2 TeV can be explained, in the nucleon effective field theory, as a destructive interference between the $V_0 V_0 \to D_3^0 +\gamma$ and
$V_0 V_0 \to (V_+ V_-)^* \to D_3^0 +\gamma$ diagrams. More details will be given in
\cite{future}.

Note that the bound state formed, $D_3^0$, is not generically the ground state; the SM electroweak
interactions, and likely also the strong interactions, favour the spin-0 singlet $D_1^0$ to be lightest isotope.
This implies that the triplet will subsequently decay to the ground
state through a magnetic transition with rate $\Gamma\sim \kappa^2
\alpha \sqrt{E_{B_1} E_{B_3}} (E_{B_1}-E_{B_3})^2/M^2 $
\cite{Redi:2018muu}, leading to a second monochromatic photon signal
with energy $E_\gamma=E_{B_1}- E_{B_3}$.
Multiple photon lines with equal rate would be a smoking gun signature
for bound
state formation in the dark sector, allowing us to easily discriminate it
from signals due to DM annihilation.

\medskip

\begin{figure*}[t]
\centering
\includegraphics[width=.48\linewidth]{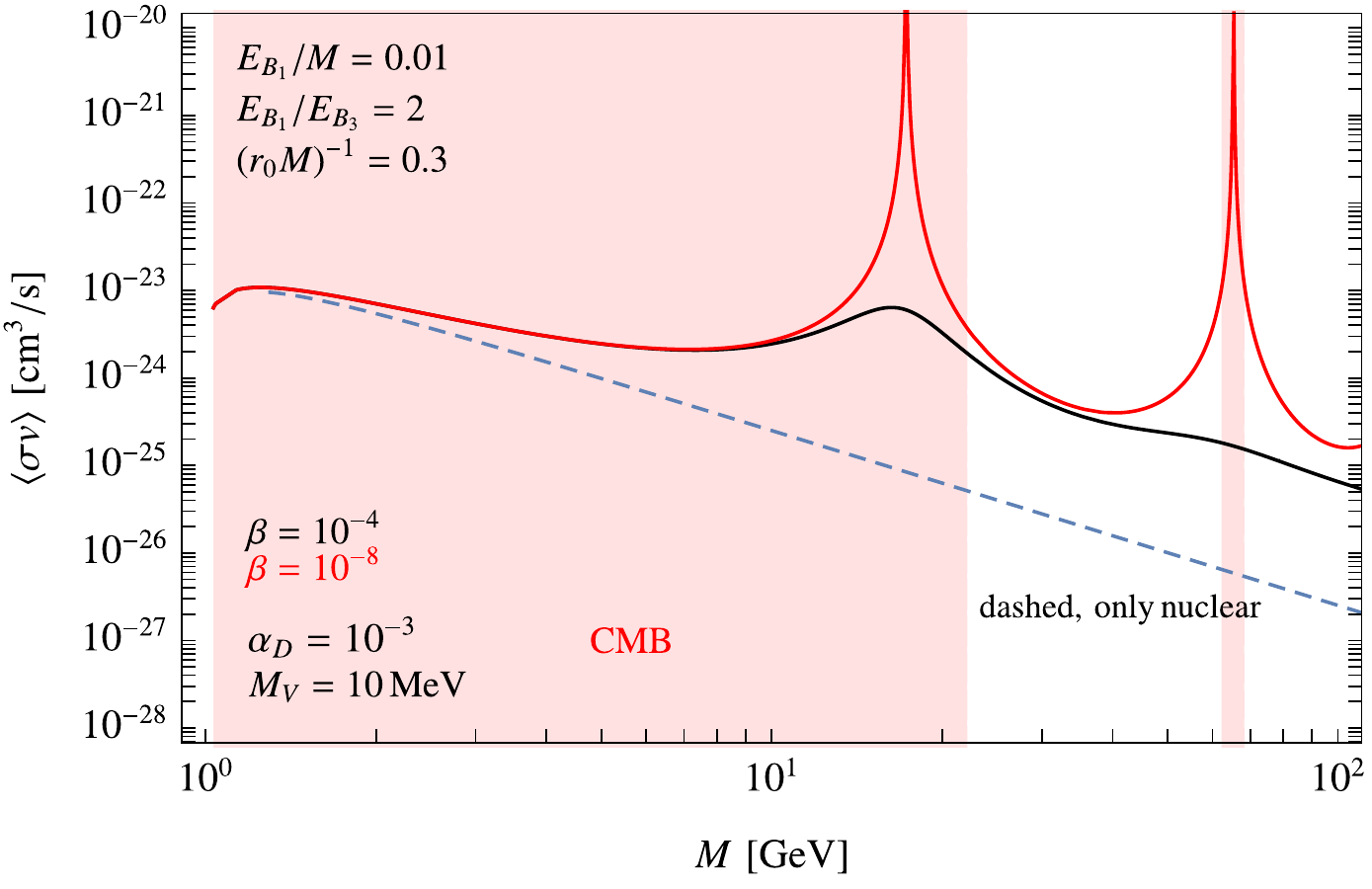}\quad 
\includegraphics[width=.48\linewidth]{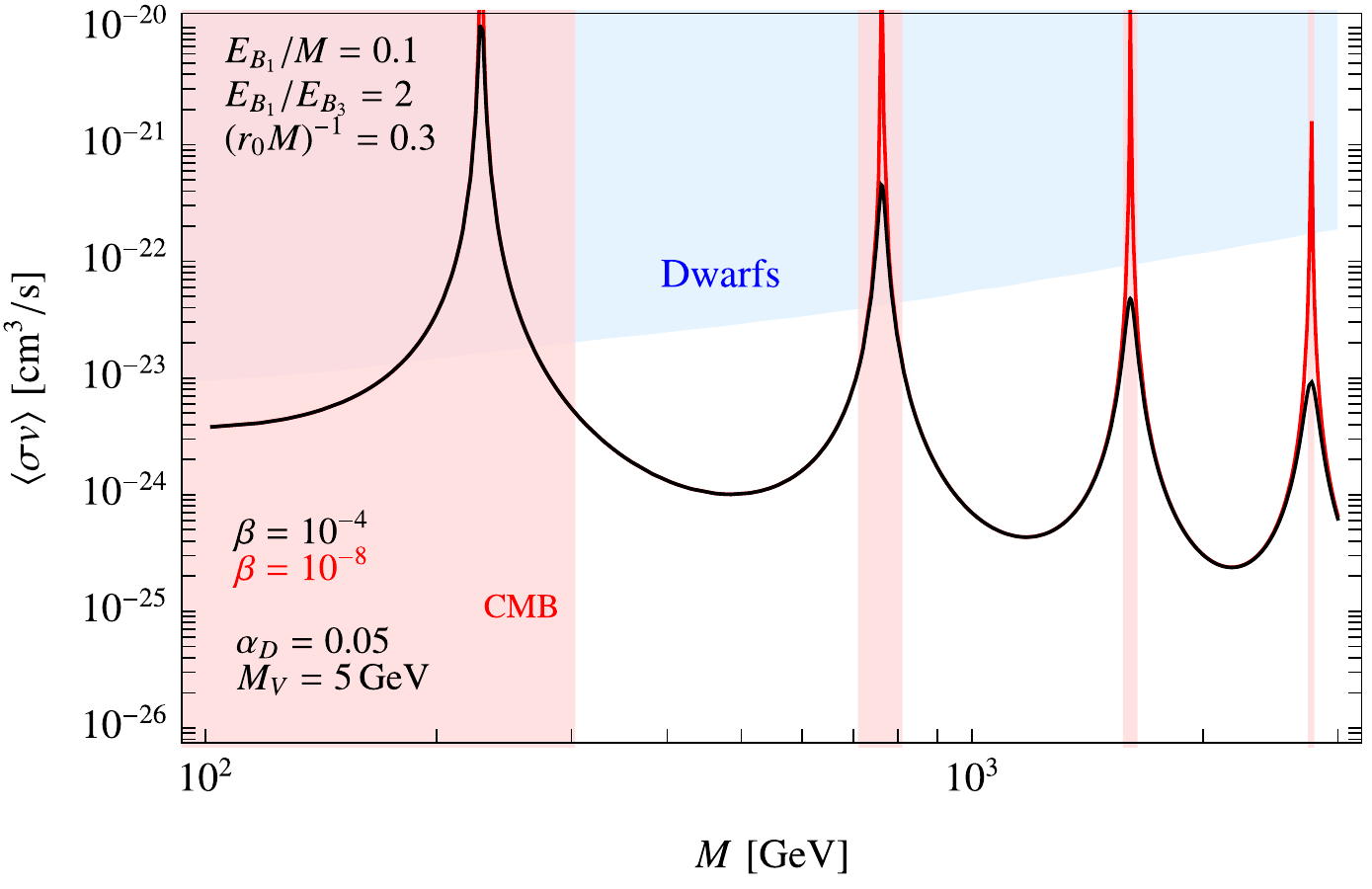}
\vspace{-0.2cm}
\caption{
\label{fig:singlet-indirect} \it
 Cross sections for dark deuterium formation $N_+ N_-\to D_1+\gamma_D$
 as a function of mass in models with SM-singlet composite DM
 and a dark photon. The light red regions are excluded by the CMB
 ($\beta\sim 10^{-8}$), see  eq.~\eqref{boundCMB}. Left panel:
 Constraints for light DM and small binding energies and couplings, as
 dictated by the CMB constraints. Right panel: Constraints for
 intermediate DM masses and sizeable binding energies and couplings;
 indirect detection constraints (light blue region) arise from diffuse
 $\gamma$-ray spectra from dwarf spheroidal galaxies as extracted from
 \cite{Profumo:2017obk} for 10\% branching
 fraction for $\gamma_D\to\tau\tau$ \cite{Cirelli:2016rnw}).
}
\end{figure*}

\paragraph{\bf Composite SM-singlet DM}\,\,
In models with SM-singlet nucleons, formation of heavier nuclei
requires interaction with a (SM-neutral) light state to carry away the
binding energy emitted in the process. This state can be a dark photon.
We assume that DM belongs to an asymmetric dark baryon doublet, $(N_+,\,N_-)$ with
equal mass  and opposite dark photon charge. This can be realised with
a confining SU(3) dark gauge theory with 2
degenerate flavours with opposite unit charges. More general assignments are possible.

The nuclear potential has a flavour-singlet spin-1 channel, and a flavour-triplet spin-0 channel. 
Including the dark photon interaction, the potential in the neutral sector is described by 
\be
V_{Q=0}^{S=0,1}= - \alpha_D \frac {e^{-M_{V} r}}r + V^N_{3,1}(r)\,.
\ee
We neglect charged channels, which are repulsive and lead to smaller cross sections.

Through a magnetic transition $N_+$ and $N_-$ can form bound states
with spin-0 (1), with the emission of a dark photon with energy
$E_{B_3}$ ($E_{B_1}$) respectively.  The
cross section for the process $N_+ N_- \to D_{1,3} + \gamma_D$ can be
computed using (\ref{eq:numerical}) with an extra factor of 1/4 to
account for distinguishable particles in the initial state.  The dark photon then decays to the SM through kinetic mixing between
the dark and hypercharge field strengths,
\begin{equation}
\mathscr{L}_D= -\frac 1 4 F_{D\,\mu\nu}F_D^{\mu\nu}- \frac 1 2 M_{V}^2 V_\mu V^\mu -\frac {\epsilon}{2 c_W} F_{D\,\mu\nu}B^{\mu\nu}\,.
\end{equation}
The dark photon phenomenology is similar to that of weakly coupled models,
see for example \cite{Cirelli:2016rnw} for the allowed region of the ($\epsilon, M_V$) parameter space.

In order to estimate the indirect detection bounds due to diffuse
photons we recast the analysis of \cite{Profumo:2017obk}.  
If DM is light, bound state formation is also strongly limited by
hydrogen re-ionization, since the CMB provides a rather
model-independent bound on the energy injected into the thermal photon bath
after recombination \cite{Slatyer:2009yq}. By using the result from
PLANCK \cite{Ade:2015xua}, this bound translates in our case to
\begin{equation}\label{boundCMB}
\langle \sigma_{D\gamma^*} v_{\rm rel}\rangle_{\rm CMB}< \frac {8.2 \times 10^{-28}\, {\rm cm^3}{s^{-1}}}{f_{\rm eff}}\times  \frac {M^2}{E_B^2}\times  \frac {E_B}{\rm GeV}\,,
\end{equation}
where the efficiency factor $f_{\rm eff}$ depends mildly on the decay
channel. For our purposes we will take $f_{\rm eff}\approx
0.5$. Neglecting long-distance effects (that tend to increase the
cross section), the nuclear rate in eq.~\eqref{magnetic-xsec} indicates
that binding energies have to be relatively small in order to satisfy
the strong CMB bounds, namely $E_{B}/M\lesssim 10^{-3} (M/\GeV)^{6/5}
(0.001/\alpha_D)^{2/5}$.

Estimates of the bounds on this scenario from dwarf spheroidal galaxies
and the CMB are shown in Fig.~\ref{fig:singlet-indirect} with the
latter always yielding the stronger constraint.  The
strongest bound arises from the formation of $D_1$ which is enhanced
with respect to $D_3$ production by a relative factor of
$(E_{B_1}/E_{B_3})^{5/2}$. The limit from
  dwarf spheroidals
is sensitive to the dark photon branching fraction, and could vary by
a factor of 5 in either direction for a different choice of mass.
For our specific choice of parameters we have verified that the cosmological
production of dark deuterium is small, see \cite{Redi:2018muu}.
Changing the parameters may result in dark deuterium, and possibly heavier dark
nuclei, being synthesised primordially.  This would reduce the indirect-detection
rate, but could also potentially trigger processes such as dark
tritium formation. Indeed, barring bottle-necks, light nuclear DM
could produce a sizeable population of heavy dark nuclei
\cite{Krnjaic:2014xza,Hardy:2014mqa} resulting in novel
phenomena that would merit detailed
analysis.

\begin{figure}[t!]
\centering
\includegraphics[width=.96\linewidth]{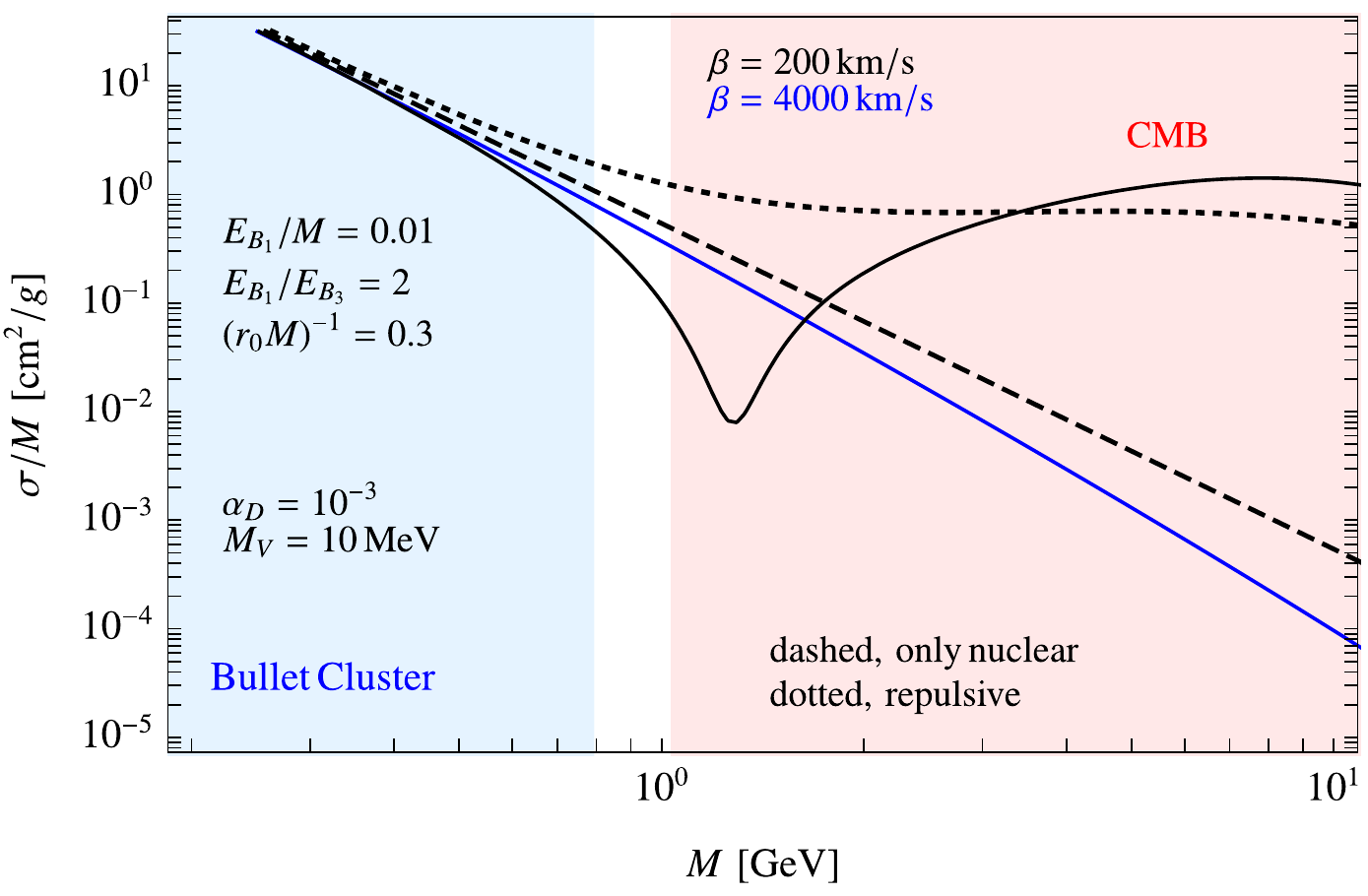}
\vspace{-0.2cm}
\caption{
\label{fig:SIDM}\it
Elastic cross section from DM self-interactions as a function of DM
mass, for the allowed region of Fig.~\ref{fig:singlet-indirect}
(left). The bullet cluster constraint corresponds to $\beta\approx
4000$ km/s, while $\beta\approx 200$km/s is a velocity typically found
in galactic cores \cite{Tulin:2017ara}.} 
\end{figure}

Our framework will also be constrained by the
  size of DM self-interactions.  These are particularly important for binding energies smaller than the mediator mass, where bound state formation by dark photon emission is
kinematically forbidden and the strong CMB bound no longer holds.  The $s$-wave elastic cross section is given by  $\sigma_{\rm el} = 4\pi/p^2 \sin^2\delta\approx 4\pi a^2/(1+p^2 a^2)$. 
We compute it by extracting phase
shifts from solutions of the Schroedinger equation eq.~\eqref{equation} using $e^{2i\delta}=e^{-i p
  r_\infty}(u'(r_\infty)+ i p u(r_\infty))/\sqrt{4\pi}$ ; in different
regions of parameter space these can be dominated either by the
nuclear forces or by the long range interactions. 
The cross section is strongly dependent on the character of the force
mediated by the massive dark photon. For opposite-sign DM particles
the  potential is attractive and the rate displays resonance peaks where the cross section saturates to $\sigma_{\rm
  el}\sim 4\pi/p^2$. For
same-sign particles there is no resonance structure but the enhancement can still be
sizeable. We illustrate this phenomenon in Fig. \ref{fig:SIDM} where the vector
boson mass has been tuned to yield a small region of dark matter
parameter space which is consistent with both CMB and bullet cluster
constraints \cite{Tulin:2017ara}. Decreasing the binding energy relative to the mediator
mass would open up this allowed region as bound state formation
becomes kinematically forbidden.  Note that in some regions of parameter space the velocity dependence
of the cross section allows the Bullet Cluster
constraint to be satisfied, while simultaneously
giving rise at lower velocities to a cross section $\sigma_{\rm el}/M\gtrsim {\rm cm^2}/{\rm g}$
that could explain observed small-scale properties of DM, see [28] for a review.

\paragraph{\bf Summary and outlook.}\,\,
In this letter we studied the indirect detection signal associated
with the formation of bound states of DM, due to the emission of quanta with energy equal to the binding energy of the bound state. 
This can lead to a monochromatic photon line or diffuse $\gamma$-ray
emission within reach of existing experiments such as FERMI.
De-excitation to the ground state could produce additional lines;
a striking signature of bound state formation
that is easily distinguishable from annihilating DM.

This mechanism is particularly relevant for detection of
asymmetric DM, which does not annihilate and would not typically
give rise to a measurable indirect-detection signal. 
It is also relevant for thermal DM, where the photons emitted in bound
state formation would be complementary to the signal from direct
annihilation of DM to photons.

We focused on magnetic dipole interactions of (dark) nuclear DM in
 two simple and compelling scenarios.  Despite the strongly-coupled
 nature of the nuclear interactions the production cross section for
 bound states can be calculated at leading
 order in terms of the binding energies.  Determining these in
 a strongly-coupled SU($N$) gauge theory is an interesting problem
 that merits further study, and could be solved on the lattice.

In this letter we have just skimmed the surface of the fascinating
phenomenology of strongly-coupled dark matter bound states.  Similar effects
can arise due to electric dipole interactions for example, or emission of `dark pions'.  Furthermore for large binding energies
emission of $W$ and $Z$ bosons may become kinematically allowed, leading to novel signatures.


\medskip
\paragraph{ \it Acknowledgements.}\,\,
{\small We thank Filippo Sala for useful comments. This work is supported by MIUR grants PRIN 2017FMJFMW and 2017L5W2PT, Ente Cassa di Risparmio di Firenze
and INFN grant STRONG.}


\bibliography{biblio}

\end{document}